\documentclass[aps, prx, twocolumn, amssymb, amsmath, superscriptaddress]{revtex4-1}

\usepackage{bm}
\usepackage{times}
\usepackage{graphicx}
\usepackage[colorlinks,citecolor=blue,linkcolor=blue]{hyperref}
\usepackage{array}
\usepackage{float}
\usepackage{colortbl}
\usepackage{multirow}
\usepackage{tabularx}

\begin{document}

\title{Effects of layer stacking and strain on electronic transport in 2D tin monoxide}
\author{Yanfeng Ge}
\affiliation{State Key Laboratory of Metastable Materials Science and Technology \& Key Laboratory for Microstructural Material Physics of Hebei Province, School of Science, Yanshan University, Qinhuangdao, 066004, China}


\author{Yong Liu}\email{yongliu@ysu.edu.cn}
\affiliation{State Key Laboratory of Metastable Materials Science and Technology \& Key Laboratory for Microstructural Material Physics of Hebei Province, School of Science, Yanshan University, Qinhuangdao, 066004, China}
\date{\today}

\begin{abstract}
Tin monoxide is an interesting two-dimensional material because of the rare oxide semiconductor with bipolar conductivity. However, the lower room temperature mobility limits the applications of SnO in the future. Thus, we systematically investigate the effects of the different layer structures and strains on the electron-phonon coupling and phonon-limited mobility of SnO. The A$_{2u}$ phonon mode in the high frequency region is the main contributor coupling  with electron for the different layer structures. And the orbital hybridization of Sn atoms existing only in bilayer structure changes the conduction band edge and decreases the electron-phonon coupling conspicuously, thus the electronic transport performance of bilayer is superior to others. In addition, the compressive strain of $\epsilon$=-1.0\% in monolayer structure makes CBM consist of two valleys at $\Gamma$ point and along M-$\Gamma$ line, also leads to the intervalley electronic scattering assisted by E$_g$-1 mode. However, the electron-phonon coupling regional transferring from high frequency (A$_{2u}$) to low frequency (E$_g$-1) results in the little significant change of mobility.

\end{abstract}

\maketitle

\section{Introduction}

It is well known that the dimension is one of crucial roles on the properties of the materials because the dimensionality reduction can give rise to the significant change. Thus the last few years have witnessed an explosion of interest in developing two-dimensional (2D) materials and understanding their properties since the discovery of graphene~\cite{Novoselov2004}. A common characteristic of 2D materials is the weak van der Waals interaction between the layers and strong bonding within the layers. Beyond the graphene, 2D materials include Xenes~\cite{Molle2017,Mannix2017}, metal chalcogenides~\cite{Chhowalla2013,Li2016,Zhou2016}, MXenes~\cite{Liu2016,Anasori2017}, layered oxide materials~\cite{Gupta2015,Mas-Balleste2011} and others, have received considerable attentions. The most monumental reason is that the many motivating properties only exists in the 2D materials. For example, the suitable bandgap, ultrahigh carrier mobility and other novel quantum effects make them apply to the electronic, optical and logic devices~\cite{Radisavljevic2011,Geim2013,Saji2016,Xiao2012}.

Among the 2D materials, IV-VI compounds provide an opportunity for sustainable electronic and photonic systems recently due to the various structures, earth-abundant and nontoxic characteristics~\cite{Zhou2016,Zhou20161}.
The ultrathin IV-VI compounds (SnS~\cite{De2013,Song2013,Huang20141,Su2015}, SnSe~\cite{Su2013,Zhou20151}, SnSSe~\cite{Pan2013}) have been mechanically exfoliated and applied in high performance field effect transistors.
In addition, IV-VI compounds have demonstrated many interesting physical properties to be worthy of an in-depth study~\cite{Kamal2016}. For example, the direct bandgap of $\sim$1.5 eV make SnS have a high absorption coefficient and to be a promising candidate for solar cells, photodetectors and photocatalytic water splitting~\cite{Sinsermsuksakul2012,Xia2016,Chowdhury2017}. The transformation from the amorphous to the crystalline state of SnSe under the laser shows the advantage over the memory devices~\cite{Micoulaut2008,Wang2011,Chung2008}. Theoretical works also predict that SnSe has layer dependent bandgap and can transit from an indirect to direct bandgap semiconductor when the thickness decreases to monolayer, indicating the potential applications in optical and optoelectronic devices~\cite{Huang2014}. Furthermore, the multiferroic~\cite{Wang2017,Wu2016}, piezoelectricity~\cite{Fei2015,Gomes2015} and topological insulator~\cite{Hsieh2012,Yang2014,Liu2015} are also predicted in the family of IV-VI compounds and greatly extend their application field in the future.

\begin{figure}[htp!]
\centerline{\includegraphics[width=0.5\textwidth]{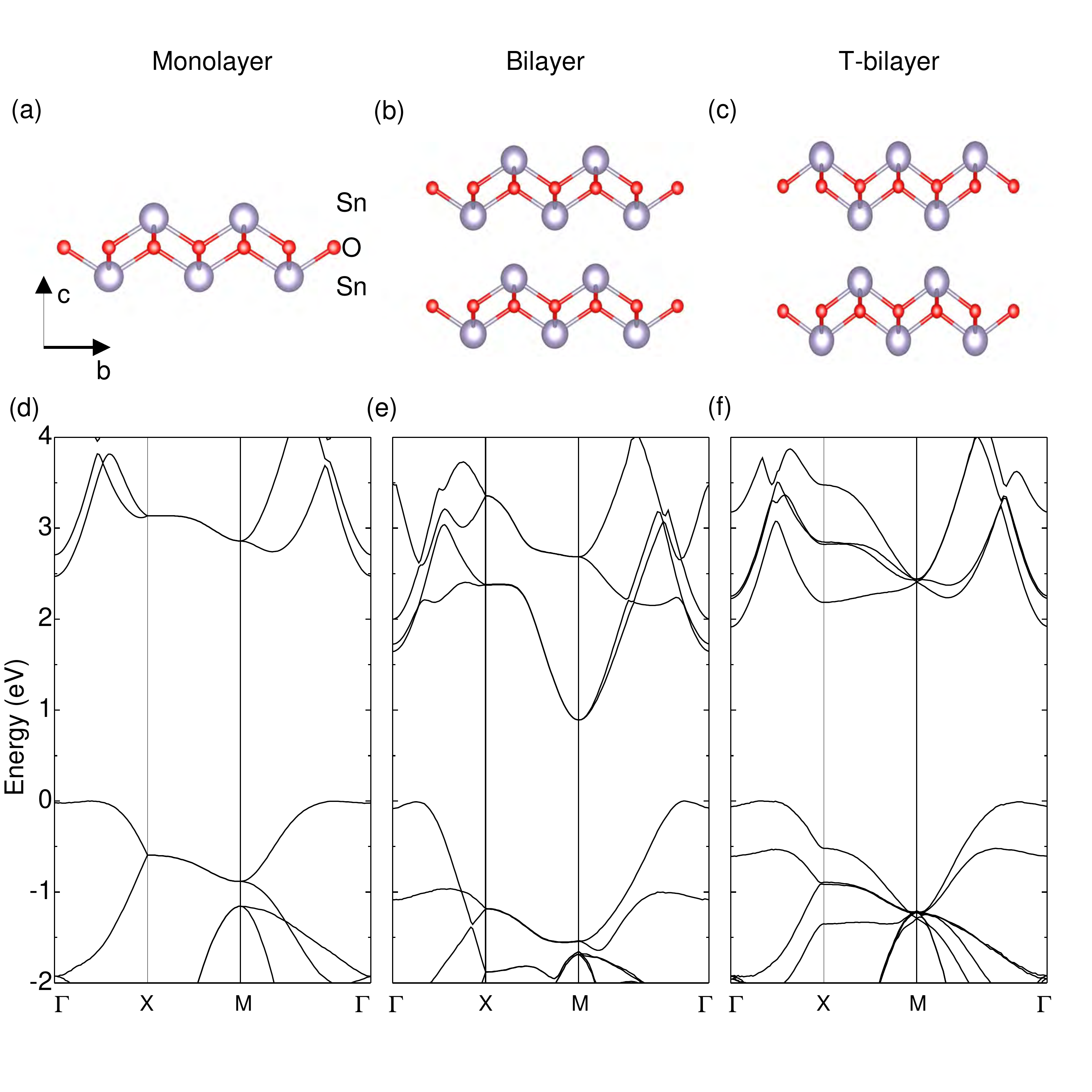}}
\caption{{\bf Lattice structures and electronic band structures of different layer structures.} The lattice structures of (a) monolayer, (b) bilayer and (c) T-bilayer. The band structures of (d) monolayer, (e) bilayer and (f) T-bilayer. The monolayer structure has a direct bandgap of $\sim$ 2.5 eV at $\Gamma$ point. The both types of bilayer have indirect bandgap, which are $\sim$ 1.0 eV and $\sim$ 1.9 eV, respectively. The VBM and CBM of T-bilayer structure locate around $\Gamma$ point. However, the CBM of bilayer structure around M point is distinctly different from other two case.
\label{fig:band}}
\end{figure}

Tin monoxide (SnO) is an interesting semiconductor and promising for a wide variety of technological applications~\cite{Seixas2016}. The stable phase of SnO has a tetragonal crystallographic structure (P4/nmm space group~\cite{Watson2001,Izumi1981,Moreno1994}). The specific lone pair~\cite{Zhou2015} of 5s electrons in SnO leads to the dipole-dipole interaction and the unique structure versus other IV-VI compounds, which makes SnO to be a rare oxide semiconductor with bipolar conductivity~\cite{Hosono2011,Varley2013,Caraveo-Frescas2013}. The van der Waals interaction along the [001] crystal direction also make SnO form a layered structure with Sn-O-Sn sequence~\cite{Lefebvre1998,Pannetier1980,Pan2001}, as shown in Fig.~\ref{fig:band}. Due to a large direct optical bandgap~\cite{Togo2006,Ogo2008,Liang2010,Quackenbush2013}, the possible coexistence of electrical conductivity and optical transparency make it idea for the invisible electronic devices~\cite{Thomas1997}.
As an essential physical properties of the application potential in multifunctional electronic devices, electronic transport of SnO has been reported by many experimental study. Field effect transistors using SnO have been developed~\cite{Ou2008,Lee2010,Nomura2011,Saji2016,Qiang2017} and show P-type conduction with the room temperature mobility $\sim$ 1 cm$^{2}$/(V$\cdot$s), much lower than that of MoS$_2$~\cite{Radisavljevic2011,Wang2012,Das2013} and phosphorene~\cite{Li2014}, and also limits the use of SnO in the future. The improvement of the mobility by the executable experimental method becomes a problem demanding prompt solution.

In the experiments, several kinds of preparation methods~\cite{He2016} can control production of high-quality 2D materials with a selected number of layers. It is found that many physical properties, especially the band edge structure, depend on the number of layers strongly. And the twist angle between adjacent layer~\cite{Chen2016,Xin2016,Cao20181,Cao20182} can also affect electron structure. Moreover, Strain engineering~\cite{Bissett2014} has been successfully used to tailor the properties of 2D materials~\cite{Wu2017,Duerloo2014,Rodin2014,Zhu2015}, such as crystal structure, bandgap, phonon and so on.

Considering the great influence of electron-phonon coupling on the electronic transport of materials, such as MoS$_2$~\cite{Ge2014} and phosphorene~\cite{Qiao2015}, here we study the electron-phonon coupling and phonon-limited mobility of SnO modulated by the different layer structures and strains based on first-principles calculations with Boltzmann transport theory. It is found that A$_{2u}$ phonon mode in the high frequency region contributes the major electron-phonon coupling for the different layer structures. It is noteworthy that the significant change of conduction band edge of bilayer results in the great decrease of electron-phonon coupling so can improve the electronic transport performance. In the monolayer structure, the compressive strain of $\epsilon$=-1.0\% though impacts the conduction band edge greatly and change the region of main electron-phonon coupling in phonon spectrum, its comprehensive effect on the mobility is minimal, when the other strains also have little effect on the electronic transport.

\section{Methods}

Our calculation is based on the semiclassical Boltzmann transport theory. The transport electron-phonon coupling constant $\lambda_{tr}$ can be obtained by,
\begin{eqnarray}
\lambda_{tr}=2\int^{\infty}_{0}\omega^{-1}\alpha^2_{tr}\rm{F}(\omega)\rm{d}\omega,
\label{eq:lambda}
\end{eqnarray}
where $\alpha^2_{tr}\rm{F}(\omega)$ is transport spectral function~\cite{allen1978}. The relaxation time $\tau$ and the temperature dependence of mobility ${\mu}(T)$ can be expressed as,
\begin{eqnarray}
\begin{aligned}
\tau^{-1}=
(\frac{\textstyle4\pi{k_B}T}{\hbar})
\int\frac{d\omega}{\omega}\frac{\tilde{\omega}^2}{\sinh^2\tilde{\omega}}\alpha^2_{tr}\rm{F}(\omega),
\end{aligned}
\label{eq:tau}
\end{eqnarray}

\begin{eqnarray}
\mu(T)=\frac{2e {\emph N_ \emph F} \langle{v^2_x}\rangle }{n_{\rm 2D}\textstyle{S_{cell}}} \tau,
\end{eqnarray}
where $\langle{v^2_x}\rangle$ is the average square of the Fermi velocity along the $x$ direction, $S_{cell}$ is the area of unit cell.

Technical details of the calculations are as follows. All calculations, including the electronic structures, the phonon spectra, and the electron-phonon couplings, were carried out using the ABINIT package~\cite{Gonze19971,Gonze19972,Gonze2005,Gonze2009} with the local-density approximation (LDA). The ion and electron interactions are treated with the Hartwigsen-Goedecker-Hutter (HGH) pseudopotentials~\cite{Hartwigsen1998}.
The strain was introduced by adjusted the lattice constant $a$ of the monolayer SnO with the strain capacity $\varepsilon=(a-a_0)/a_0\times100\%$.
By requiring convergence of results, the kinetic energy cutoff of $600$~eV and the Monkhorst-Pack $k$-mesh of 30$\times$30$\times$1 were used in all calculations about the electronic ground-state properties. The phonon spectra and the electron-phonon couplings were calculated on a 15$\times$15$\times$1 $q$-grid using the density functional perturbation theory (DFPT)~\cite{Baroni2001}. Because of the semiconductive property of SnO, carrier doping was necessary for the study of electronic transport properties and we only considered electron doping with doping concentration n$_{2D}$ = 0.5$\times$10$^{13}$ cm$^{-2}$. It is a reasonable value for the experimental doping technology and make Fermi level locate around the conduction band edge.

\section{Results}

\subsection{Number of layers and stacking effects}

The monolayer SnO consists of three atomic layers, where oxygen layer is sandwiched between two tin layers [Fig.~\ref{fig:band}(a)]. And the equilibrium lattice constant is found to be a$_0$=b$_0$=3.835 \AA. The corresponding band structure shows the direct bandgap of $\sim$ 2.5 eV with the valence band maximum (VBM) and conduction band minimum (CBM) both locating around $\Gamma$ point [Fig.~\ref{fig:band}(a)]. And the band structure nearby VBM is approximate local flat band which causes in the heavy electronic effective mass.

\begin{figure}[htp!]
\centerline{\includegraphics[width=0.5\textwidth]{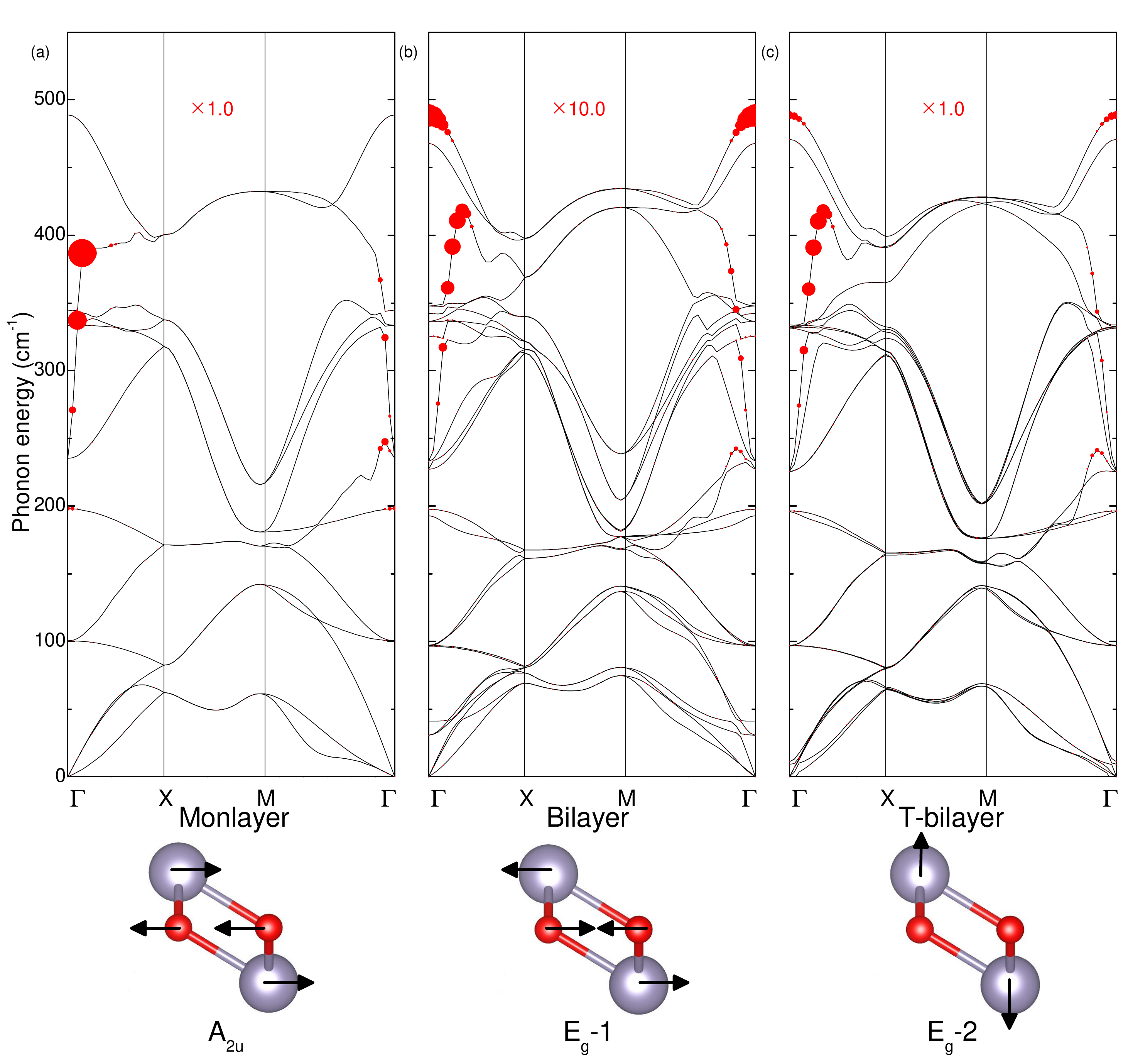}}
\caption{{\bf Phonon spectra and phonon linewidths of different layer structures.} Phonon spectra and phonon linewidths of (d) monolayer, (e) bilayer and (f) T-bilayer. The magnitude of the phonon linewidth is indicated by the size of the red error bar, and the magnitude for Bilayer is plotted with ten times the real values. The large phonon linewidths mainly focus on $\Gamma$ point in the high frequency region (350$\sim$400 cm$^{-1}$).
\label{fig:phonon}}
\end{figure}

Furthermore, the number of layers and stacking types are also considered in the present work.
First, the prototypical bilayer structure (denoted by bilayer for simplicity) is similar to the bulk structure (P4/nmm) but has a vacuum layer of 16 \AA\ in order to be 2D structure [Fig.~\ref{fig:band}(b)]. The space between two adjacent Sn in the different layers is 2.67 \AA\ along the z axis and locate the midpoint of a and b lattice vectors, respectively. The band structure of bilayer show that CBM change from $\Gamma$ point (Sn:p$_y$ orbit) to M point (Sn:s+p$_z$ orbits)~\cite{Zhou2015} and valence band around $\Gamma$ point (Sn:p$_y$ orbit) has Mexican-hat-like band structure [Fig.~\ref{fig:band}(e)]. The indirect bandgap of $\sim$ 1.0 eV in bilayer is much narrower than that of monolayer, which is obviously different from the other 2D materials~\cite{Chen2016,Xin2016}. The important cause is the interlayer Sn-Sn interactions~\cite{Zhou2015}, although the existence of weak van der Waals interactions between adjacent layers. The weak corresponding orbital hybridization of Sn atoms in the different layer leads to the bonding-antibonding splitting with inversely proportional to the space between two Sn atoms. Second, one of the bilayer is translated one-half lattice constant along b axis to constitute the new bilayer structure (denoted by T-bilayer for simplicity), as shown in Fig.~\ref{fig:band}(c). The space between two adjacent Sn in the different layers is 4.05 \AA\ along the z axis, larger than that of bilayer, and results in smaller bonding-antibonding splitting. Thus, the band structure of T-bilayer is similar to the monolayer with a smaller bandgap of $\sim$ 1.9 eV [Fig.~\ref{fig:band}(f)].
And the Mexican-hat-like band or flat band around VBM in three structures gives rise to the low P-type carrier mobility, so the N-type carrier in only considered in the present work.

\begin{figure}[htp!]
\centerline{\includegraphics[width=0.5\textwidth]{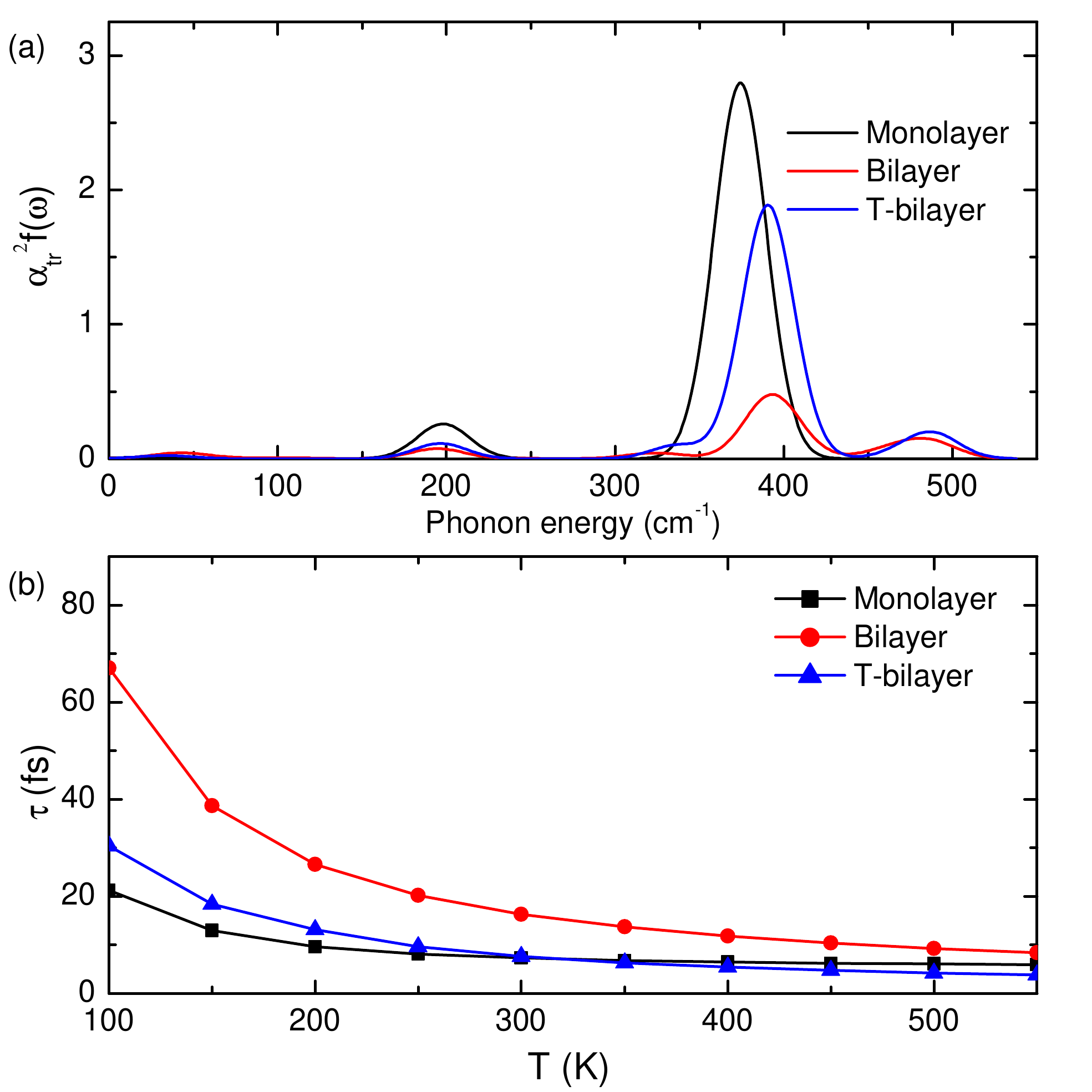}}
\caption{{\bf Transport spectral function and relaxation time of different layer structures.} (a) Transport spectral functions $\alpha_{tr}$ of monolayer, bilayer and T-bilayer. The main peak appears around $\sim$ 380 cm$^{-1}$ in the high frequency region, consistent with the results of phonon linewidths. And the peak value of Bilayer is much lower than those of other two cases. (b) The electronic relaxation time $\tau$ of monolayer, bilayer and T-bilayer as the function of temperature T.
\label{fig:afw}}
\end{figure}

By using the DFPT, we have calculated the phonon spectra and the electron-phonon couplings of monolayer, bilayer and T-bilayer. As shown in Fig.~\ref{fig:phonon}, the absence of imaginary frequency in the phonon spectra ensures the dynamic stabilities of three cases. According to the group theory, the following five irreducible representations at the $\Gamma$ point are denoted the optic vibrational modes: $\Gamma=A_{1g}+B_{1g}+2E_{g}+A_{2u}+E_{u}$~\cite{Saji2016} and Fig.~\ref{fig:phonon}(d) shows the three main optical vibration modes ($A_{2u}$, $E_{g}$-1 and $E_{g}$-2) coupling with electron strongly, as demonstrated in the following discussions. The magnitude of the phonon linewidth is indicated by the size of the red error bar in Fig.~\ref{fig:phonon}. It is found that the high frequency phonon in the range of 350$\sim$400 cm$^{-1}$ has the largest phonon linewidth in the monmolayer SnO, corresponding to the relative vibration between Sn sublattice and O sublattice in the xy plane (irreducible representation: A$_{2u}$). With regard to the bilayer, due to the significant changes of conduction band, the values of phonon linewidth decrease markedly, as shown in Fig.~\ref{fig:phonon}(b). And phonon linewidth of T-bilayer is slightly smaller than that of monolayer for the analogical band structures.

\begin{figure}[htp!]
\centerline{\includegraphics[width=0.5\textwidth]{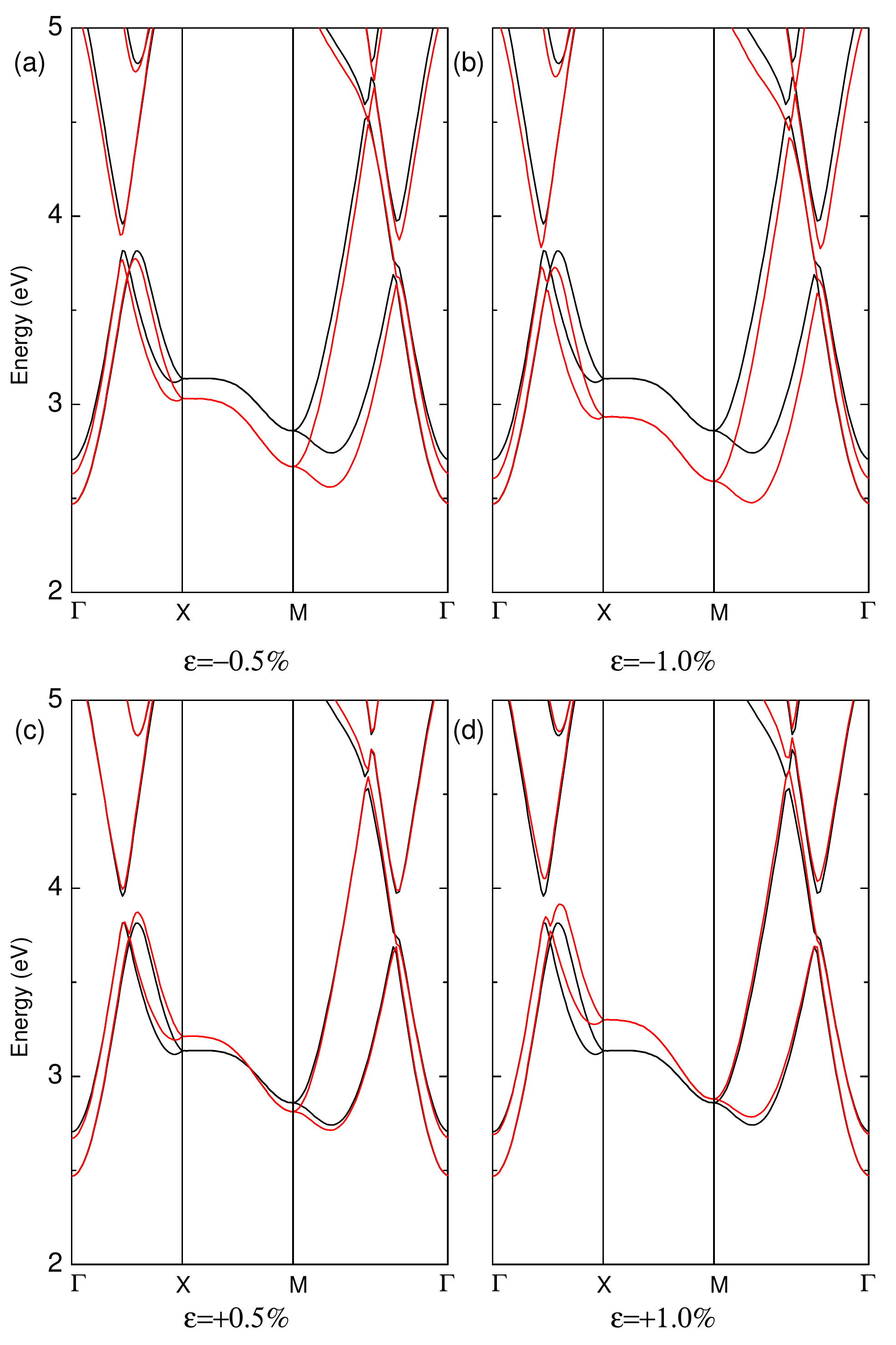}}
\caption{{\bf Electronic band structures of different strains.} The band structures under (a,b) compression strains ($\epsilon$=-0.5\%,-1.0\%) and (c,d) tensile strains ($\epsilon$=0.5\%,1.0\%). The black line indicates the band structure of strain-free monolayer SnO. The compression strains mainly influence the conduction band around X and M points as well as the X points for the case of tensile strains.
\label{fig:s-band}}
\end{figure}

As shown in Fig.~\ref{fig:afw}(a), The main peaks in the transport spectral function $\alpha_{tr}$ of monolayer also demonstrate that the strong electron-phonon coupling derive from the phonon in the range of 350$\sim$400 cm$^{-1}$, in accordance with the above results of phonon linewidths, so the case of T-bilayer. In contrast, the much smaller phonon linewidths of bilayer generate the lower peak in the overall transport spectral function. According to the Eq~\ref{eq:tau}, the electronic relaxation times $\tau$ of three cases are shown in Fig.~\ref{fig:afw}(b). The monolayer and T-bilayer have close results for the similar band structure, phonon spectra and phonon linewidths. Two cases both have $\tau$ of $\sim$10 fs at room temperature. More than that, the weaker electron-phonon coupling of bilayer give rise to the much longer electronic relax time than those of others, such as $\tau$=20 fs at room temperature.

\begin{figure}[htp!]
\centerline{\includegraphics[width=0.5\textwidth]{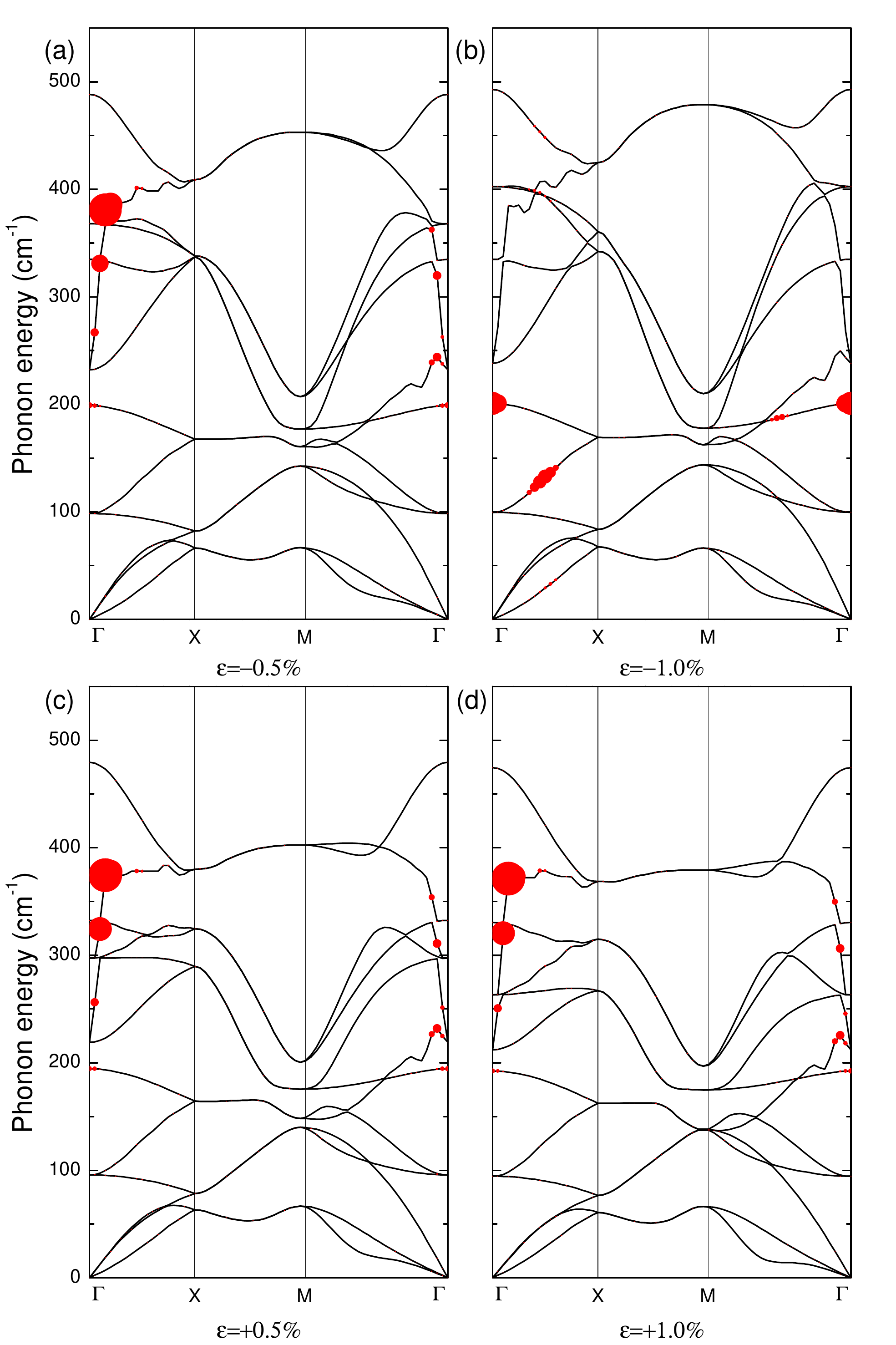}}
\caption{{\bf Phonon spectra and phonon linewidths of different strains.} Phonon spectra and phonon linewidths under (a,b) compression strains ($\epsilon$=-0.5\%,-1.0\%) and (c,d) tensile strains ($\epsilon$=0.5\%,1.0\%). The large phonon linewidths mainly focus on $\Gamma$ point in the high frequency region (350$\sim$400 cm$^{-1}$) except for $\epsilon$=-1.0\%, which has large phonon linewidths at $\Gamma$ point and midpoint of $\Gamma$-X line in the range of 150$\sim$200 cm$^{-1}$.
\label{fig:s-phonon}}
\end{figure}

\subsection{Strain effect}

Besides the effects of number of layers and stacking types, we also study the influences of different extents and types of strains in the monolayer structure, including tensile and compressive strains. To ensure the dynamics stability of lattice structures, the strains no more than 1\% are only considered in the present work, discussed later. Under the compressive strains, the conduction bands around X and M points are gradually approaching the CBM with the increase of strain, as shown in Fig.~\ref{fig:s-band}. Especially, compressive strain of $\epsilon$=-1.0\% make the CBM consist of two parts, the valleys at $\Gamma$ point and along M-$\Gamma$ line. However, the tensile strains almost little impact on the CBM except the slight rise of conduction bands around X point.

\begin{figure}[htp!]
\centerline{\includegraphics[width=0.5\textwidth]{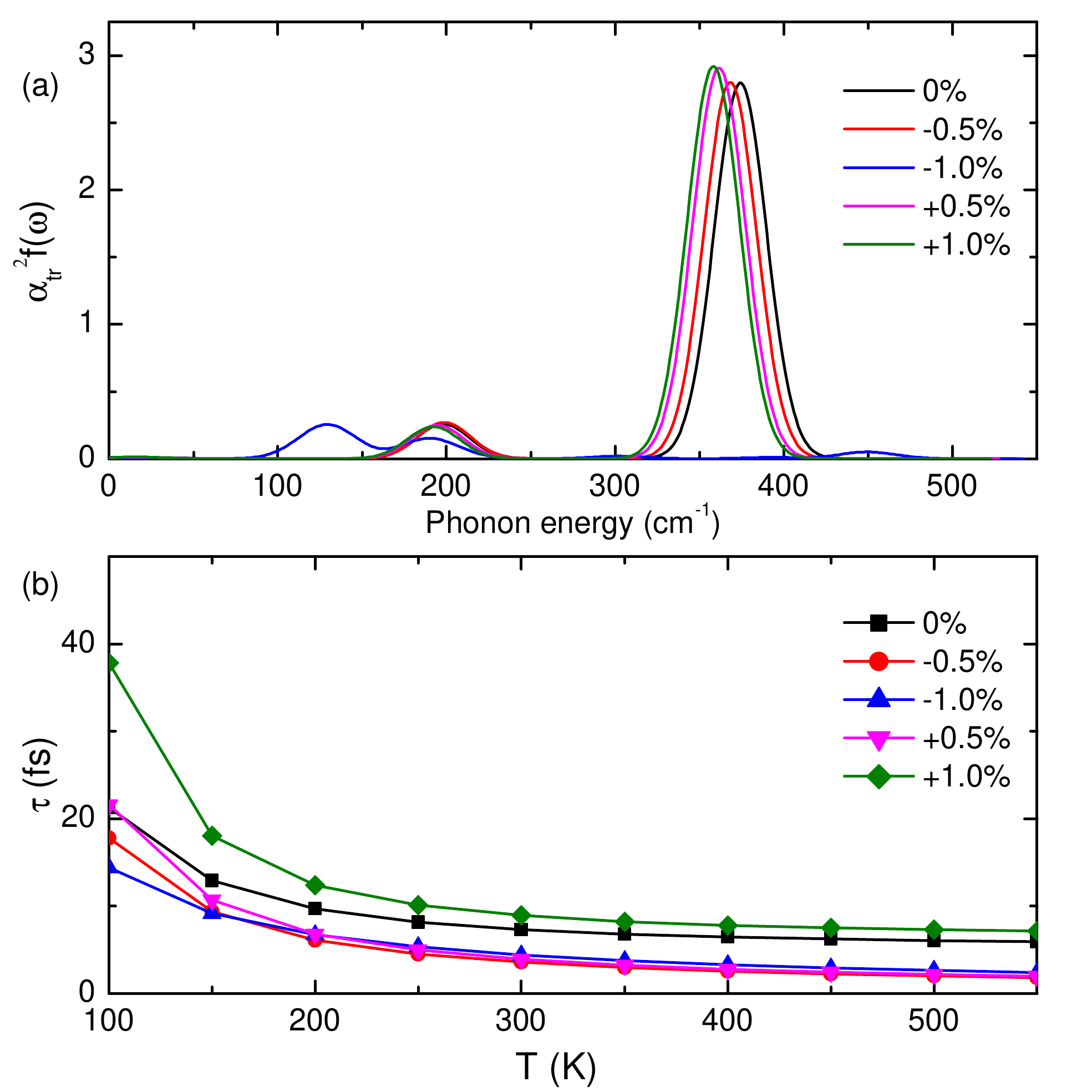}}
\caption{{\bf Transport spectral function and relaxation time of different strains.} (a) Transport spectral functions $\alpha_{tr}$ and (b) the electronic relaxation time $\tau$ under compression strains ($\epsilon$=-0.5\%,-1.0\%) and tensile strains ($\epsilon$=0.5\%,1.0\%). The main peak appears around $\sim$ 380 cm$^{-1}$ in the high frequency region except for $\epsilon$=-1.0\%.
\label{fig:s-afw}}
\end{figure}

In the phonon spectra of different extents and types of strains [Fig.~\ref{fig:s-phonon}], it is found that the compressive strains increase the atomic vibration frequencies due to reducing the distance between atoms and strengthening the bond energy, just opposite to the tensile strains. Because two tensile strains and compressive strain of $\epsilon$=-0.5\% are almost no influence on conduction band edge, the results about phonon linewidths show the similar little effects of strains. For the case of $\epsilon$=-1.0\%, the region of large phonon linewidths occur in the medium frequency and mainly contains two part of compositions, which are at $\Gamma$ point and midpoint of $\Gamma$-X line, respectively. Two phonon modes both belong to the irreducible representations of E$_g$. One is the inplane vibrations of Sn and O atoms within their sublattice and with respect to sublattice of each other ($\sim$120 cm$^{-1}$), as E$_g$-1 shown in Fig.~\ref{fig:phonon}. The other one is out-of-plane vibrations of two Sn atoms ($\sim$200 cm$^{-1}$), as E$_g$-2 shown in Fig.~\ref{fig:phonon}. And the secondary peak around $\sim$200 cm$^{-1}$ of monolayer [Fig.~\ref{fig:afw}(a)] shows that E$_g$-2 mode is also weak coupling with electron in the strain-free condition. The major reason for the change of electron-phonon coupling is the effect of $\epsilon$=-1.0\% on the conduction band edge, in especial around M point. And the E$_g$-1 mode at midpoint of $\Gamma$-X line assist the intervalley electronic scattering between the valleys around M point. On the opposite, the phonon modes around $\Gamma$ point assist the intravalley electronic scattering.

The transport spectral function $\alpha_{tr}$ under different strains also illustrate the significant change of electron-phonon coupling, as shown in Fig.~\ref{fig:s-afw}(a). For the case of $\epsilon$=-1.0\%, the main peak in the high frequency has a sharp decrease. The new peak around $\sim$120 cm$^{-1}$ results from the contribution of E$_g$-1 mode coupling with electrons. According to Eq.~\ref{eq:lambda}, the small peak in the low frequency region still produces large electron-phonon coupling. Therefore, The relaxation time $\tau$ of $\epsilon$=-1.0\% is lower than the prototypical monolayer. Finally, the strains considered in the present work have no remarkable influence on electronic relaxation time, as shown in Fig.~\ref{fig:s-afw}(b).

\begin{figure}[htp!]
\centerline{\includegraphics[width=0.5\textwidth]{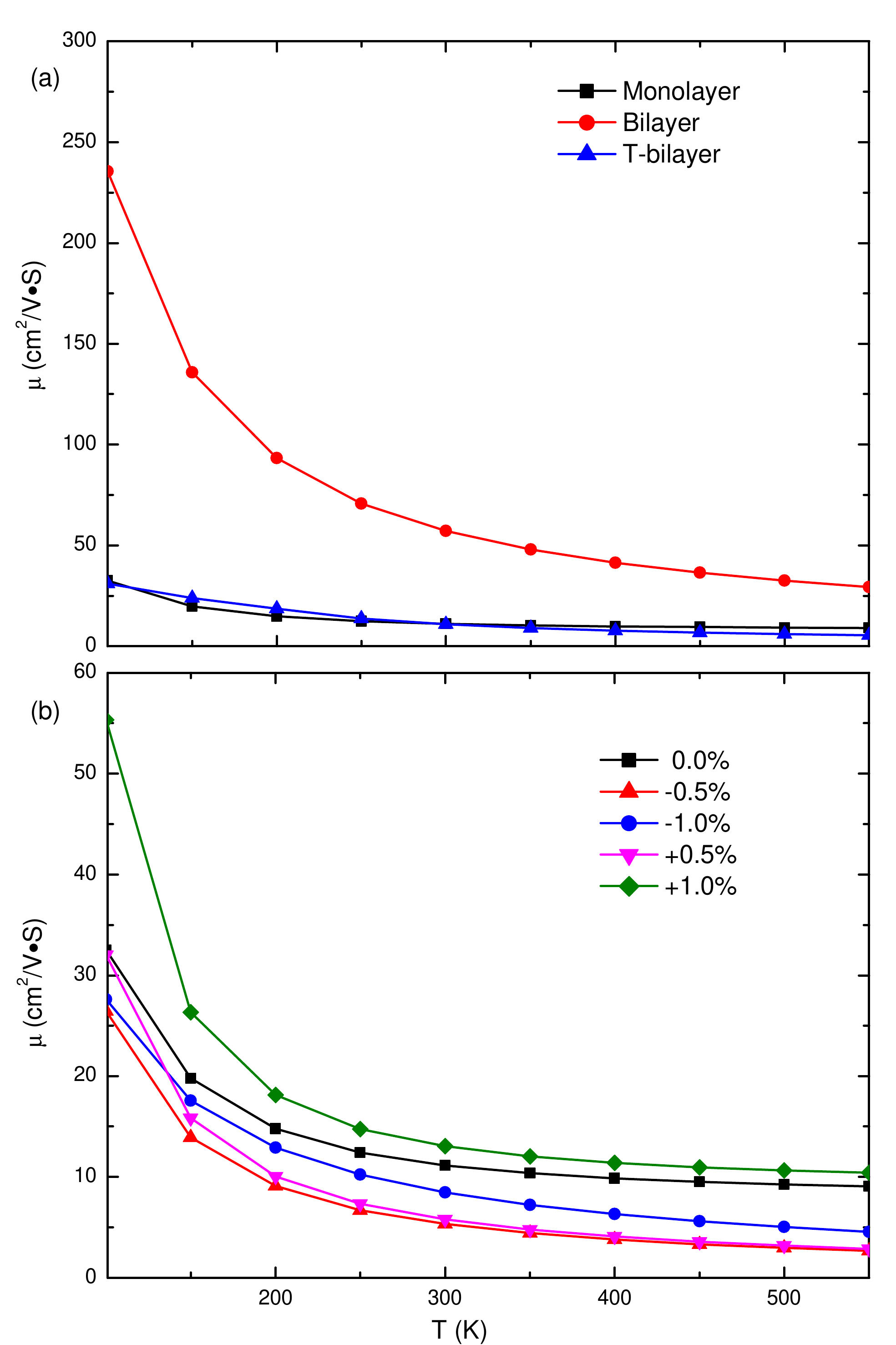}}
\caption{{\bf Mobility of SnO.} Mobility$\mu$ as a function of the temperature T (a) of monolayer, bilayer and T-bilayer and (b) monolayer with various strains ($\epsilon$=-0.5\%,-1.0\%,0.5\%,1.0\%).
\label{fig:mobility}}
\end{figure}

\subsection{Mobility}

The carrier mobility $\mu$ as a function of the temperature for the structures with different layers and various strains are plotted in Fig.~\ref{fig:mobility}. Firstly, $\mu$ of bilayer is much higher than those of monolayer and T-bilayer, due to the prominent change of conduction band edge as well as the electron-phonon coupling. At room temperature, bilayer has $\mu$=76 cm$^{2}$/(V$\cdot$s) with six times the mobility of monolayer or T-bilayer, and the enhancement of bilayer is more significant when the temperature drops [Fig.~\ref{fig:mobility}(a)], such as 230 cm$^{2}$/(V$\cdot$s) at 100 K. Secondly, the results of strains show that there is no notable difference between the carrier mobility under various strains [Fig.~\ref{fig:mobility}(b)]. At room temperature, $\mu$ for all cases has the low value of $\sim$10 cm$^{2}$/(V$\cdot$s).

\section{Summary}

In summary, we have studied the effect of the number of layer and strains on the phonon-limited mobility of SnO.
In the strain-free condition, it is found that the coupling of electron with A$_{2u}$ phonon mode in the high frequency region is the strongest for three types of layer structures. And the interaction between Sn atoms from different layers in bilayer structure changes the CBM obviously. Hence the bilayer has highest mobility in three cases. After introduction of strain in monolayer structure, the compressive strain of $\epsilon$=-1.0\% leads to that CBM consists of two valleys at $\Gamma$ point and along M-$\Gamma$ line, So the intervalley electronic scattering assisted by E$_g$-1 mode only appears in this case,  But the electron-phonon coupling regional transferring from high frequency (A$_{2u}$) to low frequency (E$_g$-1) results in the little significant change of electronic transport, which is also present in other strains. This study provides the fundamental information about the electron-phonon coupling and electronic transport property for deeper research works.

\begin{acknowledgments}
This work was supported by the NSFC (Grants No.11747054), the Specialized Research Fund for the Doctoral Program of Higher Education of China (Grant No.2018M631760), the Project of Heibei Educational Department, China (No. ZD2018015 and QN2018012), and the Advanced Postdoctoral Programs of Hebei Province (No.B2017003004).
\end{acknowledgments}

\end{document}